\documentclass[12pt, draftclsnofoot, onecolumn]{IEEEtran}
\usepackage{amsmath}
\usepackage{amssymb}
\usepackage{amsthm}
\usepackage{array}
\usepackage{fixmath}
\usepackage[final]{graphicx}
\usepackage{cite}
\usepackage{color}
\usepackage{changes}
\usepackage{algorithm}
\usepackage{algorithmic}
\usepackage{hhline}
\usepackage{float}
\usepackage{acronym}
\restylefloat{table}

\DeclareMathAlphabet{\mathbit}{OML}{cmr}{bx}{it}

\newcommand{\B}[1]{\mathbit{#1}}

\DeclareMathOperator{\Transpose}{T}
\DeclareMathOperator{\Hermitian}{H}
\newcommand{\Tr}{{\Transpose}}
\newcommand{\He}{{\Hermitian}}

\DeclareMathOperator{\blockdiag}{blockdiag}

\DeclareMathOperator{\rank}{rank}

\newcommand{\LBS}{{N_\text{RF}}}
\newcommand{\Lrk}{{R_{\text{RF}}}}

\newcommand{\BS}{{\text{BS}}}
\newcommand{\MS}{{\text{MS}}}

\newcommand{\eul}{{\text{e}}}
\newcommand{\imj}{{\text{j}}}
\newcommand{\Npk}{{N_{\text{p},k}}}

\DeclareMathOperator{\diag}{diag}

\DeclareMathOperator*{\argmax}{argmax}

\title{Hybrid LISA for Wideband Multiuser Millimeter Wave Communication Systems under Beam Squint}

\author{Jos\'e~P.~Gonz\'alez-Coma, ~\IEEEmembership{Member,~IEEE,}
	Wolfgang~Utschick,~\IEEEmembership{Senior Member,~IEEE,}
	and~Luis~Castedo,~\IEEEmembership{Senior Member,~IEEE}
	\thanks{Jos\'e~P.~Gonz\'alez-Coma~and~Luis~Castedo are with the University of A Coru\~na, CITIC, Spain. e-mail: \{jose.gcoma, luis\}@udc.es}
	\thanks{Wolfgang~Utschick is with Professur Methoden der Signalverarbeitung, Technische Universit\"at M\"unchen, Germany. e-mail: utschick@tum.de}
	\thanks{This work has been funded by Xunta de Galicia (ED431C 2016- 045, ED341D R2016/012, ED431G/01), AEI of Spain (TEC2015-69648-REDC, TEC2016-75067-C4-1-R), and ERDF funds (AEI/FEDER, EU).}}

\begin{document}
\acrodef{ADC}[ADC]{Analog-to-Digital Converter}
\acrodef{AO}[AO]{Alternating Optimization}
\acrodef{AoA}[AoA]{Angle of Arrival}
\acrodef{AoD}[AoD]{Angle of Departure}
\acrodef{AWGN}[AWGN]{Additive White Gaussian Noise}
\acrodef{BS}[BS]{Base Station}
\acrodef{MIMO}[MIMO]{Multiple-Input Multiple-Output}
\acrodef{ASK}[ASK]{Amplitude-Shift Keying}
\acrodef{BER}[BER]{Bit Error Ratio}
\acrodef{BF}[BF]{Beamforming}
\acrodef{BFN}[BFN]{Beamforming Network}
\acrodef{BST}[BST]{Barium–Strontium–Titanate}
\acrodef{CDF}[CDF]{cumulative distribution function}
\acrodef{CMOS}[CMOS]{Complementary Metal–Oxide–Semiconductor}
\acrodef{DAC}[DAC]{Digital to Analog Converter}
\acrodef{DCS}[DCS]{Digital Communication System}
\acrodef{DOA}[DOA]{Direction Of Arrival}
\acrodef{DC}[DC]{Direct Current}
\acrodef{DSB}[DSB]{Double Side Band}
\acrodef{DSP}[DSP]{Digital Signal Processor}
\acrodef{ESD}[ESD]{Energy Spectral Density}
\acrodef{FoM}[FoM]{Figure of Merit}
\acrodef{FSK}[FSK]{Frequency-Shift Keying}
\acrodef{FT}[FT]{Fourier Transform}
\acrodef{HT}[HT]{Hilbert Transform}
\acrodef{ICI}[ICI]{Inter-carrier Interference}
\acrodef{IL}[IL]{Insertion Losses}
\acrodef{ISI}[ISI]{Inter-Symbol Interference}
\acrodef{LBFN}[LBFN]{Linear Beamforming Network}
\acrodef{LC}[LC]{Liquid Crystal}
\acrodef{LISA}[LISA]{Linear Successive Allocation}
\acrodef{LMD}[LMD]{Linearly Modulated Digital}
\acrodef{LOS}[LOS]{Line-of-sight}
\acrodef{MISO}[MISO]{Multiple-Input Single-Output}
\acrodef{MMIC}[MMIC]{Monolithic Microwave Integrated Circuit}
\acrodef{MEMS}[MEMS]{Micro-Electro-Mechanical Systems}
\acrodef{MRC}[MRC]{Maximum Ratio Combining}
\acrodef{mmWave}[mmWave]{millimeter wave}
\acrodef{MSLL}[MSLL]{Maximum Side-Lobe Level}
\acrodef{MS}[MS]{Mobile Station}
\acrodef{NLOS}[NLOS]{Non Line-of-sight}
\acrodef{NMLW}[NMLW]{Normalized Main-Lobe Width}
\acrodef{NPD}[NPD]{Normalized Power Density}
\acrodef{OFDM}[OFDM]{ Orthogonal frequency-division multiplexing}
\acrodef{PA}[PA]{Power Amplifier}
\acrodef{PS}[PS]{Phase Shifter}
\acrodef{pHEMT}[pHEMT]{pseudomorphic High Electron Mobility Transistor}
\acrodef{PSK}[PSK]{Phase-Shift Keying}
\acrodef{QAM}[QAM]{Quadrature Amplitude Modulation}
\acrodef{RF}[RF]{Radio Frequency}
\acrodef{RFC}[RFC]{Rayleigh Fading Channel}
\acrodef{RPDC}[RPDC]{Reconfigurable Power Divider/Combiner}
\acrodef{RTPS}[RTPS]{Reflection Type Phase Shifters}
\acrodef{SA}[SA]{Simulated Annealing}
\acrodef{SDMA}[SDMA]{space-division multiple access}
\acrodef{SER}[SER]{Symbol Error Rate}
\acrodef{SLL}[SLL]{Side-Lobe Level}
\acrodef{SR}[SR]{Sideband Radiation}
\acrodef{SNR}[SNR]{Signal-to-Noise Ratio}
\acrodef{SP3T}[SP3T]{Single-Pole Triple-Throw}
\acrodef{SSB}[SSB]{Single Side Band}
\acrodef{USB}[USB]{Upper Side Band}
\acrodef{LSB}[LSB]{Lower Side Band}
\acrodef{SWC}[SWC]{Sum-of-Weighted-Cosine}
\acrodef{SPMT}[SPMT]{single-pole-multiple-throw}
\acrodef{SPST}[SPST]{single-pole-single-throw}
\acrodef{STA}[STA]{Static Array}
\acrodef{TM}[TM]{Time Modulation}
\acrodef{TMA}[TMA]{Time-Modulated Array}
\acrodef{ULA}[ULA]{Uniform Linear Array}
\acrodef{VGA}[VGA]{Variable Gain Amplifier}


\maketitle
\begin{abstract}
	This work jointly addresses user scheduling and precoder/combiner design in the downlink of a wideband \ac{mmWave} communications system.  We consider \ac{OFDM} modulation to overcome channel frequency selectivity and obtain a number of equivalent narrowband channels. Hence, the main challenge is that the analog preprocessing network is frequency flat and common to all the users at the transmitter side. Moreover, the effect of the signal bandwidth over the \ac{ULA} steering vectors has to be taken into account to design the hybrid precoders and combiners. The proposed algorithmic solution is based on \ac{LISA}, which greedily allocates streams to different users and computes the corresponding precoders and combiners. By taking into account the rank limitations imposed by the hardware at transmission and reception, the performance loss in terms of achievable sum rate for the hybrid approach is negligible. Numerical experiments show that the proposed method exhibits excellent performance with reasonable computational complexity.      
\end{abstract}

\section{Introduction}
One of the key features of \ac{mmWave} is the availability of large bandwidths, making this technology a promising candidate to satisfy the increasing capacity demand of the fifth generation of cellular networks. To avoid the high costs in terms of powers consumption and implementation, hybrid analog and digital architectures have been proposed \cite{AlMoGoHe14}. These architectures exhibit clear advantages in terms of costs, although they lack of the flexibility of the purely digital solutions. Indeed, the design of hybrid precoders and combiners has recently drawn a lot of attention, especially in multiuser scenarios (see e.g. \cite{ZhNgYu17,UtStJoLu18,DaCl17}). Despite the variety of methods proposed in the literature to determine the precoders and combiners, these designs cannot be directly applied to the wideband scenario. The reason behind this claim is that the analog precoder has to be jointly designed for all users and subcarriers.

Several hybrid precoding designs have been proposed for wideband transmissions, most of them focusing on the single user case \cite{AlHe16,YuShZhLe16,IwVuCaLuXuUt16,SoYu17,BuAnFoUgCo18}. Multiuser settings were addressed in  \cite{BoLeHaVa16,YuZhLe2017Alternating} but at the expense of more complex schemes including two (or more) \ac{PS} for each connection between a \ac{RF} chain and a single antenna element. Authors in \cite{ChWaLi17} provide a method showing good performance, but its practical use is limited since it is restricted to the transmission of a single spatial stream. Likewise, a single stream per-user is allocated following the strategy provided in \cite{KwChSu17} for the multiuser \ac{MISO} case. A more general setup was considered in \cite{GoRoGoCaHe18}, where the hybrid digital and analog precoders and combiners leverage the common structure of the channel response matrices among different subcarriers \cite{VeGoHe17}.

However, the common structural assumption is not accurate except in narrowband situations with low relative signal bandwidth with respect to the carrier frequency. Moreover, in the general case, the steering vectors are affected by the beam squint effect, that is, a change in the beam direction which depends on the signal bandwidth.    Additionally, prior work on wideband \ac{mmWave} systems, to the best of authors knowledge, does not consider the impact on precoding and combining designs caused by the signal bandwidth except for \cite{CaLaHo17}. Therein, authors consider a fixed analog codebook as well as the array gain losses caused by beam squint. Such assumptions are taken into account while allocating the power among the different subcarriers. Nevertheless, the design of hybrid analog and digital precoders and combiners was not addressed in \cite{CaLaHo17}.     

\subsection{Contribution}

We propose an algorithmic solution based on Hybrid-LISA \cite{UtStJoLu18} that accounts for the user scheduling together with the design of hybrid analog and digital precoders and combiners. Although a common structure of the channel for different subcarriers is not assumed, some similarity of the column and vector spaces even holds when considering beam squint. In the following, this similarity is exploited to find the frequency flat analog precoders and combiners. The proposed iterative method allocates an additional stream such that interference to previously allocated streams is suppressed, jointly selecting the best user candidate for all subcarriers. The corresponding precoder and combiner are then computed according to this choice. Finally, the digital precoders are calculated to remove the remaining interference. Furthermore, the baseband stages of the hybrid precoders and combiners provide the flexibility of switching off some subcarriers for a particular stream. 

We also analyze the computational complexity of the proposed algorithm, and pose an alternative approach to reduce the complexity of the search performed at each iteration. This approach is based on dividing the signal bandwidth in different subbands, and using a representative subcarrier for each of them in the greedy selection process. 

Numerical experiments are provided to reveal the performance of the proposed method. Moreover, we show the impact of the beam squint effect over the state of the art wideband hybrid precoding solutions. The results show that the common structural assumption of \cite{VeGoHe17} is not accurate in general, and that the consequent impact on the system performance is significant. Also, we numerically evaluate the average equivalent channel gains for each subcarrier under the zero-forcing constraint. According to the results obtained with the proposed method, increasing the signal bandwidth for a given carrier frequency leads to smaller equivalent channel gains for the edge frequencies, thus confirming that the beam squint effect reduces the achievable sum rate compared to the scenario assuming a common frequency flat channel structure.  

\section{System Model}
The system setup considered in this work consists of a \ac{BS} (transmitter) equipped with $N$ transmit antennas and $K$ \ac{MS} (users) with $R$ antennas each. The \ac{BS} can allocate zero or more streams to the users, with the maximum number of streams, $N_\text{s}$, limited by the number of \ac{RF} chains at the \ac{BS}, $\LBS$. On the other side of the communication link, the number of \ac{RF} chains at each user, $\Lrk$, restricts the number of independent data streams allocated to such user. 
Due to the large bandwidth available in \ac{mmWave} systems, the wideband signals pass through a frequency selective channel. To combat this effect, we consider \ac{OFDM} symbols with cyclic prefix long enough to avoid \ac{ICI}. 
The data is linearly processed in two stages with the baseband precoder $\B{P}_{\text{D},k}[\ell]\in\mathbb{C}^{\LBS\times N_{\text{s},k}}$ at subcarrier $ \ell $, followed by the frequency flat analog precoder $\B{P}_\text{A}\in\mathbb{C}^{N\times \LBS}$. At the user's end, the received signal is linearly filtered with the analog and baseband combiners, i.e., frequency flat $\B{G}_\text{A}\in\mathbb{C}^{R\times \Lrk}$ and frequency selective $\B{G}_{\text{D},k}[\ell]\in\mathbb{C}^{\Lrk\times N_{\text{s},k}}$. Since the RF filters are implemented using analog
phase shifters, its entries are restricted to a constant modulus $|[\B{P}_\text{A}]_{i,j}|^2=|[\B{G}_\text{A}]_{m,n}|^2=1$.  
\subsection{Channel Model}
\label{subsec:channel}
In this work, we focus on the scenario where both the \ac{BS} and the users are equipped with \acp{ULA}, for simplicity. The steering vectors at the \ac{BS} are then given by \cite{Va02}
\begin{equation}
\B{a}_\BS(\phi)[\ell]=\frac{1}{\sqrt{N}}[1 ,\eul^{\imj 2\pi\mathsf{k}[\ell]\mathsf{d}\sin(\phi)},\ldots,\eul^{\imj 2\pi\mathsf{k}[\ell]\mathsf{d}\sin(\phi)(N-1)}]	
\label{eq:steering_vector}
\end{equation}
where $N$ is the number of transmit antennas, $\mathsf{d}$ is the inter-element spacing, $\phi$ is the \ac{AoD}, and $\mathsf{k}[\ell]=\frac{1}{\lambda[\ell]}=\frac{f[\ell]}{c}$ corresponds to the wavenumber. We assume a subcarrier dependent wavenumber as a consequence of modeling the transmitted passband signal carrier frequency as $f[\ell]=f_c+\xi[\ell]$, where $f_c$ is the central frequency and $\xi[\ell]$ represents a frequency offset. Consider now the signal bandwidth is $B$ and the number of subcarriers for the \ac{OFDM} modulation is $L$. Accordingly, the frequency interval between subcarriers is $\frac{B}{L}$ and $\xi[\ell]=(\ell-\frac{L+1}{2})\frac{B}{L}$. By defining the matrix $\B{\Xi}_\BS(\phi)[\ell]=\diag(1,\eul^{\imj2\pi \xi[\ell]\mathsf{d}\sin(\phi)},\ldots,\eul^{\imj2\pi \xi[\ell]\mathsf{d}\sin(\phi)(N-1)})$, \eqref{eq:steering_vector} can be rewritten as 
\begin{equation}
\B{a}_\BS(\phi)[\ell]=\B{\Xi}_\BS(\phi)[\ell]\B{a}_\BS(\phi).
\end{equation}
A similar reasoning applies for the receive steering vectors at the \ac{MS}, i.e.,
\begin{equation}
\B{a}_\MS(\theta)[\ell]=\B{\Xi}_\MS(\theta)[\ell]\B{a}_\MS(\theta)
\end{equation}
where $\theta $ denotes the \ac{AoA}. Hence, the channel response matrix for the $k$-th user at the $\ell$-th subcarrier is given by the contribution of the $\Npk$ propagation paths
\begin{align}
\B{H}_k[\ell]&=\sum_{p=1}^{\Npk}\alpha_{k,p}[\ell]\B{a}_\MS(\theta)[\ell]\B{a}_\BS^\He(\phi)[\ell]\notag\\
&=\B{A}_{\text{MS},k}[\ell]\B{\Delta}_k[\ell]\B{A}_{\BS,k}^\He[\ell]
\label{eq:channelModelFreq}
\end{align}
where the $p$-th column of $\B{A}_{\text{MS},k}[\ell]$ and $\B{A}_{\BS,k}[\ell]$ is $\B{a}_{\text{MS},k}[\ell](\theta_{k,p})$ and $\B{a}_{\text{BS}}[\ell](\phi_{k,p})$, respectively, and $\B{\Delta}_k[\ell]$ is a diagonal matrix containing the channel gains $\alpha_{k,p}[\ell]$. The $ \theta_{k,p} $ and $ \phi_{k,p} $ forms the pair of \ac{AoA} and \ac{AoD} of the corresponding frequency flat $p$-th propagation path out of a total number of $\Npk$ paths between the \ac{BS} and the $k$-th \ac{MS}. If the wavenumber is subcarrier independent, i.e., there is no beam squint effect, the $k$-th user $\ell$-th subcarrier channel matrix response can be approximated as \cite{GaDaWaCh15,VeGoHe17}
\begin{equation}
\B{H}_k[\ell]\approx\B{A}_{\text{MS},k}\B{\Delta}_k[\ell]\B{A}_{\BS,k}^\He,
\end{equation}
that is, $\B{\Xi}_{\text{MS},k}(\theta_{k,p})[\ell]=\mathbf{I}_R$ and $\B{\Xi}_{\text{BS},k}(\phi_{k,p})[\ell]=\mathbf{I}_N$.
In this work, however, the beam squint effect will be taken into account.  This is a more general approach when handling the large bandwidth signals usual in \ac{mmWave} systems.

\subsection{Performance Metric}
The utilization of \ac{OFDM} modulation and the absence of \ac{ICI} allows us to decompose the wideband channel into  $L$ equivalent narrow-band channels. Therefore, the received signal at user $k$ and subcarrier $\ell$  is given by
\begin{align}
\hat{\B{s}}_k[\ell]=\B{G}_\text{D,k}^\He[\ell]\B{G}_\text{A}^\He\B{H}_k[\ell]\left(\sum\limits_{i=1}^{K}\B{P}_\text{A}\B{P}_{\text{D},i}[\ell]\B{s}_i[\ell]+\B{n}_k[\ell]\right)
\label{eq:downshat}
\end{align}
where $\B{n}_k[\ell]\sim\mathcal{N}_\mathbb{C}(\mathbf{0},\sigma^2\mathbf{I}_{R})$ is the \ac{AWGN}. The performance metric considered for the optimization of the multiuser precoders and combiners is the achievable sum rate given by
\begin{align}
\label{eq:sum_rate}
\sum_{k=1}^KR_k & =\sum_{k=1}^K\frac{1}{L}\sum_{\ell=1}^{L}R_{k}[\ell]
\end{align}
with
\begin{align}
R_{k}[\ell] & =\log_2\det\big(
\mathbf{I}_{M_k}+\B{X}_k^{-1}[\ell]\B{G}_{\text{D},k}^\He[\ell]\B{Y}_k[\ell]\B{Y}_k[\ell]^\He\B{G}_{\text{D},k}\big),
\end{align}
the auxiliary matrix 
\begin{align}
\B{Y}_k[\ell]=\B{G}_\text{A}^\He\B{H}_k[\ell]\B{P}_\text{A}\B{P}_{\text{D},k}[\ell],
\end{align}  
and the interference plus noise matrix \begin{align}\B{X}_k[\ell] & =\sum_{i\neq k}\B{G}_{\text{D},k}^\He[\ell]\B{Y}_i[\ell]\B{Y}_i^\He[\ell]\B{G}_{\text{D},k}[\ell] \nonumber \\ & \quad+\sigma^2\B{G}_{\text{D},k}^\He[\ell]\B{G}_\text{A}^\He\B{G}_\text{A}\B{G}_{\text{D},k}[\ell].
\end{align} 
Note that we are assuming Gaussian signaling and the inter-user interference is treated as noise. 

It is clear that the maximization of \eqref{eq:sum_rate} under the restrictions imposed by the hardware is a very difficult problem. To leverage this complicated task, we propose to rely on a heuristic approach. Based on the LISA algorithm introduced in \cite{GuUtDi09}, and its suitability for hybrid precoding \cite{UtStJoLu18}, we propose an extension to consider wideband signals. To this end, we take into account that the analog precoder is common for all subcarriers and users. Similarly, the analog combiner of each user is common for all subcarriers, too. Consequently, we propose a greedy method for selecting data streams for a certain user. Once a stream is selected, power is allocated over the subcarriers of such stream. Moreover, the interference with previous allocated streams is suppressed for all subcarriers by means of a projection step.

\section{Hybrid precoder and combiner design}
\ac{LISA} successively performs user scheduling and the provision of precoders and combiners. To this end, we introduce the function $\pi(i)$, which allocates the $i$-th data stream to the corresponding user $\pi(i)$ in the respective iteration step. For the $\ell$-th subcarrier, the data symbol $t_i[\ell]$ is an element of the symbol vector $\B{s}_{\pi(i)}[\ell]$, and precoding and combining vectors $\B{p}_i[\ell]$ and ${\B{g}}_i[\ell]$ are columns of $\B{P}_\text{A}\B{P}_{\text{D},\pi(i)}[\ell]$ and $\B{G}_{\text{A},\pi(i)}\B{G}_{\text{D},\pi(i)}[\ell]$, respectively. According to \eqref{eq:downshat}, the received data symbol reads as 
\begin{align}
\label{eq:per_stream_eq}
\hat{t}_i[\ell]&={\B{g}}_i^\He[\ell]\B{H}_{\pi(i)}[\ell]\B{p}_i[\ell]t_i[\ell] \nonumber \\
&\quad+{\B{g}}_i^\He[\ell]\sum_{j\neq i}\B{H}_{\pi(i)}[\ell]\B{p}_j[\ell]t_j[\ell]+{\B{g}_i^\He[\ell]}\B{n}_{\pi(i)}[\ell].
\end{align}

In the first stage of the \ac{LISA} procedure, the auxiliary unit norm precoders ${\B{q}}_i[\ell]$ are selected from the nullspace of the effective channels obtained in preceding iterations, i.e., ${\B{q}}_i[\ell]\in\text{null}\{{\B{g}}_j[\ell]\B{H}_{\pi(j)}[\ell]\}_{j<i}$. The second stage of \ac{LISA} ensures the zero-forcing condition with $\B{p}_i[\ell]\in\text{null}\{{\B{g}}_j[\ell]\B{H}_{\pi(j)}[\ell]\}_{j\neq i}$, i.e., $\B{p}_i$ is computed as a linear transformation of ${\B{q}}_i[\ell]$ which removes the remaining interference. Thus, enabling a simple power allocation of the available transmit power according to the channel gains for each subcarrier and stream. In other words, the expression in \eqref{eq:sum_rate} at the $i$-th iteration reduces to
\begin{equation}
\label{eq:perStreamRate}
\sum_{k=1}^KR_{k}=\sum_{j=1}^i\frac{1}{L}\sum_{\ell=1}^L\log_2\left(1+\frac{\lambda_{j}^2[\ell]\gamma_{j}^2[\ell]}{\sigma^{2}}\right),
\end{equation} 
where $\lambda_{j}[\ell]=|{\B{g}}_j^\He[\ell]\B{H}_{\pi(j)}[\ell]\B{p}_j[\ell]|/\|\B{p}_j[\ell]\|_2$ and $\gamma_j[\ell]=\|\B{p}_j[\ell]\|_2$. 

At each iteration, the most promising stream is greedily selected using information from all subcarriers. In principal, the \ac{LISA} strategy could be individually applied for each subcarrier. However, this approach would not satisfy the constraint of frequency flat analog precoders and combiners, $\B{p}_i[\ell]=\B{P}_\text{A}\B{p}_{\text{D},i}[\ell]$ and $\B{g}_i[\ell]=\B{G}_\text{A}\B{g}_{\text{D},i}[\ell]$, and a respective approximation of the matrices by means of a hybrid decomposition technique  \cite{GoRoGoCaHe18} might cause a non-negligible performance loss. 
On the contrary, we propose to jointly allocate the streams for all subcarriers to facilitate the hybrid approach of the digital solution in \cite{UtStJoLu18}. To this end, the  auxiliary unit norm precoders $\B{q}_i[\ell] $ and combiners $\B{g}_i[\ell]$ are composed as $\B{q}_i[\ell]=\B{q}_i\beta_i[\ell]$ and $\B{g}_i[\ell]=\B{g}_i\beta_i[\ell]$. The frequency flat parts of the precoders and combiners are denoted by $ \B{q}_i $ and $ \B{g}_i $ and the frequency selective scalar $ \beta_i[\ell] $ indicates whether the stream is active for a particular subcarrier or not. 

The composite channel matrix of the $\ell$-th subcarrier after the $i$-th iteration, i.e., after $i$ data streams have been assigned to the selected users $ \pi(1),\dots,\pi(i)$, is
\begin{align}
	{\B{H}}_{\text{comp},i}[\ell]&=\begin{bmatrix}
	{\B{g}}_{1}^\He[\ell]\B{H}_{\pi(1)}[\ell]\\ \vdots \\ {\B{g}}_{i}^\He[\ell]\B{H}_{\pi(i)}[\ell]
	\end{bmatrix}, \label{eq:widebandLISAproduct2}
\end{align}
and the matrix of auxiliary precoders is given by 
\begin{align}
{\B{Q}}_i[\ell]&=\begin{bmatrix}
{\B{q}}_1[\ell] & \ldots & {\B{q}}_i[\ell]
\end{bmatrix}.
\label{eq:widebandLISAproduct3}
\end{align}
	
	\begin{equation}
	\begin{bmatrix}
	{\B{g}}_{1}^\He[\ell]\B{H}_{\pi(1)}[\ell]\\ \vdots \\ {\B{g}}_i^\He[\ell]\B{H}_{\pi(i)}[\ell]
	\end{bmatrix}\begin{bmatrix}
	{\B{q}}_{1}[\ell] & \ldots & {\B{q}}_i[\ell]
	\end{bmatrix}
	=\begin{bmatrix}
	{\B{g}}_1^\He[\ell]\B{H}_{\pi(1)}[\ell]{\B{q}}_1[\ell] & 0 & \ldots & 0\\ {\B{g}}_2^\He[\ell]\B{H}_{\pi(2)}[\ell]{\B{q}}_1[\ell] &  {\B{g}}_2^\He[\ell]\B{H}_{\pi(2)}[\ell]{\B{q}}_2[\ell] & \ldots & 0 \\
	\vdots &  \vdots & \ddots & 0  \\{\B{g}}_{i}^\He\B{H}_{\pi(i)}[\ell]{\B{q}}_1[\ell] & {\B{g}}_{i}^\He[\ell]\B{H}_{\pi(i)}[\ell]{\B{q}}_2[\ell] & \ldots & {\B{g}}_{i}^\He[\ell]\B{H}_{\pi(i)}[\ell]{\B{q}}_{i}
	\end{bmatrix}
	\label{eq:widebandLISAproduct}
	\end{equation}

The product of (\ref{eq:widebandLISAproduct2}) and (\ref{eq:widebandLISAproduct3}) constitutes the effective channel from the transmitter to the selected receivers at the respective subcarrier and iteration step. The resulting channel is formed by the frequency flat parts of the precoders and combiners, which are switched on or off for the respective subcarrier. 

\subsection{LISA First Stage}
\label{sec:FistStage}
The aim of the first stage of \ac{LISA} is to obtain the structure of \eqref{eq:widebandLISAproduct} for the products {${\B{H}}_{\text{comp},i}[\ell]{\B{Q}}_i[\ell]$} in all subcarriers. 
The total number of streams is smaller or equal to $\LBS$, i.e., $ \rank([{\B{Q}}_i[1],\ldots,{\B{Q}}_i[L]])\leq \LBS $. Likewise, the total number of assigned streams per user is limited by $\Lrk$. Recall that these constraints are unlikely to hold when applying \ac{LISA} individually to each subcarrier, which forces to drop some of the streams when performing the hybrid decomposition.

In order to identify the strongest transmit and receive directions jointly considering all subcarriers, we define
\begin{equation}
\label{eq:projectedChannels}
\B{H}_{k,i}=[\B{S}_{k,i}\B{H}_k[1]\B{T}_i,\ldots,\B{S}_{k,i}\B{H}_k[L]\B{T}_i]
\end{equation} 
as the channel matrix containing the projected channels for user $k$ and the $L$ subcarriers with the frequency flat orthogonal projection matrices $\B{S}_{k,i}$ and $\B{T}_i$. The projections ensure that inter-stream interference with previous allocated streams is zero. 

1) Initializing to $\B{T}_1=\mathbf{I}$, the update rule for $\B{T}_{i+1} $ reads as
\begin{equation}
\label{eq:projUpdate}
\B{T}_{i+1}=\B{T}_i\B{\Pi}_i^\perp=\prod_{j=1}^{i}\B{\Pi}_j^\perp=\mathbf{I}-\sum_{j=1}^i\B{\Pi}_j,
\end{equation}
where $\B{\Pi}_i^\perp$ denotes the orthogonal projector onto the nullspace of $\B{\Pi}_i$ and
the orthogonal projector $\B{\Pi}_i$ is chosen such that its column space is equal with the span of $\{\B{T}_i\B{H}_{\pi(i)}^\He[\ell]{\B{g}_{i}[\ell]}\}_{\ell=1}^L$, this is,
\begin{align}
	\operatorname{span}\left[\B{\Pi}_i\right] = \operatorname{span}\left[\B{T}_i\B{H}_{\pi(i)}^\He[\ell]{\B{g}_{i}[\ell]} \mid \ell=1,\dots,L\right].
	\label{eq:span_of_proj}
\end{align}
Thus, all active subcarriers are taken into account to perform the update. The scalar $\beta_i[\ell]$ provides the flexibility to decide whether the stream would be active for a particular subcarrier or not, depending on the power allocated in the second stage of the algorithm. Consequently, if power allocation at the $\ell$-th subcarrier is zero, it will not affect the update of the projector. Note that, due to the greedy nature of the algorithm, the determination of $ \beta_{1}[\ell],\dots,\beta_{i-1}[\ell] $ remains fixed at the $i$-th iteration step. Due to the construction of $\B{\Pi}_i$, it is apparent that $\B{T}_{i+1}$ fulfills the property of an orthogonal projector. An immediate consequence is that each iteration of the algorithm reduces the feasible subspace for subsequent precoding stages. However, in order to avoid excessive consumption of degrees of freedom in a single LISA step, it might be reasonable to approximate the projector $ \B{\Pi}_i $ such that it only covers the principle components of the right hand side in \eqref{eq:span_of_proj}. Consequently, then the constraints in \eqref{eq:widebandLISAproduct} can only be achieved approximately as well.

2) The projection matrices $\B{S}_{k,i}$ are introduced to avoid linear dependencies at the combiners. If more than one stream is allocated to the same user, the corresponding equalizers have to be mutually orthogonal. Otherwise, the power allocated to the linear dependent component of the equalizer vanishes, as shown in Appendix \ref{ap:Sproj}. 
In particular, the projector corresponding to the user $\pi(i)$ at the $i$-th iteration is updated as
\begin{equation}
\B{S}_{\pi(i),i+1}=\B{S}_{\pi(i),i}-{{{\B{g}}_i{\B{g}}_i^\He}},
\end{equation}
with unit norm vectors 
$\B{g}_i$ and $\B{S}_{k,1}=\mathbf{I}$. Obviously, the update of the projector is independent of the subcarrier index.

Now, given the bilinearly projected channel matrices $ \B{S}_{k,i}\B{H}_k[\ell]\B{T}_i $, we first assume a candidate equalizer $ \B{g}_{i}(k) $ for each user $k$ as the left singular vector corresponding to the largest singular value of the matrix in (\ref{eq:projectedChannels}), i.e.,
\begin{align}
\B{g}_{i}(k)&=\B{u}^\text{max}_{k,i}(\B{H}_{k,i}).
\label{eq:gaux}
\end{align} In order to determine the candidate auxiliary precoders and combiners, all subcarriers are still taken into consideration during this step, i.e., $\beta_i[\ell]=1,\,\forall\ell$. 

Hence, for the computation of the corresponding candidate auxiliary precoder $ \B{q}_{i}(k) $ for each user $k$, we define the following matrix including the obtained combiner as
\begin{equation}
\label{eq:equivFrecSelChannelMatrix}
\left(\B{g}_{i}^\He(k) \B{H}_{k,i}\right)^\text{T-block}
=\begin{bmatrix}
\B{g}_{i}^\He(k)\B{S}_{k,i}\B{H}_k[1]\B{T}_i\\\vdots\\ \B{g}_{i}^\He(k)\B{S}_{k,i}\B{H}_k[L]\B{T}_i
\end{bmatrix},
\end{equation}
where "T-block" refers to a blockwise version of the matrix transposition. The auxiliary precoder is then selected as the right singular vector associated with the largest singular value of the matrix (\ref{eq:equivFrecSelChannelMatrix}), i.e.,
\begin{align}
\B{q}_{i}(k) & =\B{v}_{k,i}^\text{max}\left(\left(\B{g}_{i}^\He(k) \B{H}_{k,i}\right)^\text{T-block}\right).
\label{eq:qaux}
\end{align} 
By defining the vector 
\begin{equation}
	\B{\mu}_{k,i}=\left(\B{g}_{i}^\He(k) \B{H}_{k,i}\right)^\text{T-block} \B{q}_{i}(k),
\end{equation}
consisting of all hypothetical channel gains for all subcarriers when deploying the obtained combining and auxiliary precoding candidates,
the user selection at the $i$-th iteration is performed by 
\begin{equation}
\pi(i)=\argmax_{k\in\{1,\ldots,K\}}\|\B{\mu}_{k,i}\|_p.
\label{userselection}
\end{equation}
In the following, we use $p=1$ for selecting the user with the largest sum of channel gains over all subcarriers.
Taking into account the result of (\ref{userselection}), we obtain the $i$-th pair of auxiliary precoder and combiner at the respective iteration step as $\B{q}_i=\B{q}_{i}(\pi(i))$ and $\B{g}_i=\B{g}_{i}(\pi(i))$. In the subsequent section, we present the second stage of \ac{LISA}, which updates the power allocation and $\beta_i[\ell]$ accordingly for all subcarriers. \\

\subsection{LISA Second Stage}
\label{sec:SecondStage}
The second stage of \ac{LISA} naturally matches with the determination of the baseband precoding part $\B{P}_{\text{D},k}[\ell]$ within the hybrid design. Thanks to the frequency selectivity nature of the baseband component, it is possible to remove the residual interference, resulting from the auxiliary precoder and combiner pairs designed at the first stage, individually for each subcarrier. Hence, we obtain the channel gains for the equivalent channels and perform the spatial-frequencial power allocation.  

Let us start by introducing the term $d_i[\ell]=\sum_{j=1}^{i-1}\beta_j[\ell]+1$, to denote the number of streams allocated to subcarrier $\ell$ at previous iterations and incremented because of the candidate of iteration $i$. Recall that, in general, $d_i[\ell]\neq d_i[l^\prime]$ for $l\neq l^\prime$. Moreover, we define the  matrix $\B{Q}_i=[\B{q}_1,\ldots,\B{q}_i]$ containing the precoders for all iterations of \ac{LISA}, and the selection matrices $\B{B}_i[\ell]\in\mathbb{C}^{i\times d_i[\ell]}$ whose columns are $\B{e}_j$ for $\beta_j[\ell]=1$, $j < i$, and the last column is $\B{e}_i$. Using these definitions, the selection matrix $ \B{B}_i[\ell] $ extracts the active rows from the composite channel matrix in (\ref{eq:widebandLISAproduct2}) and yields the reduced matrices
\begin{align}
{ {\B{H}}_{\text{comp},i}^{\text{red}}[\ell]} & {=\B{B}_i[\ell]^\Tr{\B{H}}_{\text{comp},i}[\ell]\in\mathbb{C}^{d_i[\ell]\times N},} \\ {{\B{Q}}_i^{\text{red}}[\ell]} & {=\B{Q}_i\B{B}_i[\ell]\in\mathbb{C}^{N\times d_i[\ell]}.}
\end{align}  
Further, we obtain the composite channel matrix
\begin{align}
{\B{H}}^{\text{red}}_{\text{comp},i}&=\blockdiag({\B{H}}^{\text{red}}_{\text{comp},i}[1],\ldots,{\B{H}}^{\text{red}}_{\text{comp},i}[L])
\end{align} 
and the precoding matrix 
\begin{align}
\B{Q}_i^{\text{red}}&=\blockdiag({\B{Q}}_i^{\text{red}}[1],\ldots,{\B{Q}}_i^{\text{red}}[L]).
\end{align} 
By multiplying these two  matrices, we obtain
\begin{equation}
\label{eq:Lreddefinition}
	{\B{L}_i={\B{H}}^{\text{red}}_{\text{comp},i}{\B{Q}}_i^{\text{red}}=\blockdiag({\B{L}}_i[1],\ldots,{\B{L}}_i[L]),}
\end{equation}
a block diagonal matrix of which each block ${\B{L}}_i[\ell]\in\mathbb{C}^{d_i[\ell]\times d_i[\ell]}$ with a lower triangular structure corresponds to the $\ell$-th subcarrier.

In the second stage of \ac{LISA}, we compute effective precoders ${\B{P}}_{\text{eff},i}$ removing the remaining interference. Indeed, multiplying ${\B{Q}}_i^{\text{red}}$ times ${\B{L}}^{-1}_i$ a diagonal matrix is obtained, and \eqref{eq:sum_rate} can be calculated stream-wise as in \eqref{eq:perStreamRate}. On the other hand, the columns of the resulting precoder are no longer unit norm vectors. To compensate this effect, the normalization matrix ${\B{\Lambda}_i=[\diag({\B{L}}_i^{-\He}{\B{L}}_i^{-1})]^{-1/2}}$ is introduced. Notice that ${\B{\Lambda}}_i=\blockdiag({\B{\Lambda}}_i[1],\ldots,{\B{\Lambda}}_i[L])$ contains $\sum_{\ell=1}^Ld_i[\ell]$ channel gains. That is, the channel gains related to the streams and subcarriers allocated at iteration $j$, for all $j<i$, together with the channel gains for all subcarriers at iteration $i$. Thus, the power allocation ${\B{\Gamma}_i=\blockdiag({\B{\Gamma}}_i[1],\ldots,{\B{\Gamma}}_i[L])}$ is a diagonal matrix whose entries are determined via waterfilling over the streams and subcarriers included in the diagonal of ${\B{\Lambda}}_i$. The matrix ${\B{\Gamma}}_i$ indicates the power allocated for the active streams, whereas for the streams and subcarriers not present in ${\B{\Lambda}}_i$ (or in ${\B{\Gamma}}_i$) it is thereby zero. This allocation satisfies the total power constraint $\sum_{j=1}^i\sum_{\ell=1}^L\gamma_{j}^2[\ell]\leq P_\text{tx}$. Using the power allocation ${\B{\Gamma}}_i$, we eventually compute the frequency selective effective precoder for iteration $i$ as
\begin{align}
\label{eq:blockDiagPrec}
{\B{P}}_{\text{eff},i} & =\blockdiag({\B{P}}_{\text{eff},i}[1],\ldots,{\B{P}}_{\text{eff},i}[L]) \nonumber \\ & =\B{Q}^{\text{red}}_i{\B{L}}_i^{-1}{\B{\Lambda}}_i{\B{\Gamma}}_i,
\end{align}
where ${\B{P}}_{\text{eff},i}[\ell]\in\mathbb{C}^{N\times d_i[\ell]}$.

A non-zero power allocation leads to the assignment $\beta_{i}[\ell]=1$, if $\gamma_i[\ell]>0$, and $\beta_{i}[\ell]=0$ otherwise, and determines the frequency selective precoders and combiners ${\B{q}}_i[\ell]$ and ${\B{g}}_i[\ell]$ in Sec. \ref{sec:FistStage}. 

\ac{LISA} greedily allocates streams at each iteration for a number of subcarriers, although additional streams cause a reduction of the available subspace due to the inter-stream interference removal. Correspondingly, the algorithm stops if allocating new streams does not lead to an increase in the achievable sum rate. Additional stopping and selection criteria are imposed by the hardware limitations of the hybrid architecture, that is, the number of iterations is restricted to $\LBS$. 

\subsection{Wideband H-LISA}
\label{sec:WHLISA}
In the following, we show that after a proper factorization the effective precoder in \eqref{eq:blockDiagPrec} satisfies the restriction imposed by the number of \ac{RF} chains. Nevertheless, such factorization yields an analog precoding matrix which does not satisfy the unit modulus constraint of the variable \ac{PS}. In order to derive the hybrid approach, we first rewrite the precoding matrix in \eqref{eq:blockDiagPrec} as
\begin{align}
\label{eq:overallPrec}
	\B{P}_{\text{eff},i}={\blockdiag(\B{Q}_i\B{B}_i[1]\B{\Psi}_i[1],\ldots,\B{Q}_i\B{B}_i[L]\B{\Psi}_i[L]),}
\end{align}
with $\B{\Psi}_i[\ell]={\B{L}}_i^{-1}[\ell]{\B{\Lambda}}_i[\ell]{\B{\Gamma}}_i[\ell]$. As previously stated the total number of \ac{LISA} iterations is always smaller than or equal to the number of RF chains at the transmitter, i.e., $i \leq \LBS$. Hence, the extension to the hybrid scenario is similar to that for the narrowband scenario \cite{UtStJoLu18,StUtJoLu17}. Therein, after algorithms convergence, the matrix $\B{Q}_i$ is substituted by its projection onto the feasible set for analog precoding as
\begin{equation}
\label{eq:HybidProj}
	[\B{P}_\text{A}]_{m,n}=\eul^{\imj\arg([\B{Q}_i]_{m,n})}.
\end{equation}
The product in \eqref{eq:Lreddefinition} for each subcarrier is then
\begin{align}{\B{H}}^{\text{red}}_{\text{comp},i}[\ell]\B{P}_\text{A}\B{B}_i[\ell]&={\B{\Upsilon}}_i[\ell],
\end{align}
 where the lower triangular structure $ {\B{L}}_i[\ell] $ has been altered to 
\begin{align} {\B{\Upsilon}}_i[\ell]&={\B{L}}_i[\ell]\B{B}_i[\ell]^\He\B{Q}_i^\He\B{P}_\text{A}\B{B}_i[\ell].\end{align}
Then, to remove the resulting inter-stream interference, we need to multiply $\B{P}_\text{A}\B{B}_i[\ell]$ times ${\B{\Upsilon}}_i^{-1}[\ell]$, similarly to the second stage of \ac{LISA} described in Sec. \ref{sec:SecondStage}. Yet again, we normalize the columns of $\B{P}_\text{A}\B{B}_i[\ell]{\B{\Upsilon}_i^{-1}[\ell]}$ by means of the product with the diagonal matrix  ${\B{\Lambda}}_i[\ell]=[\diag({\B{\Upsilon}}_i^{-\He}[\ell]{\B{\Upsilon}}_i^{-1}[\ell])]^{-1/2}$, which contains the channel gains employed to find the associated power allocation ${\B{\Gamma}}_i[\ell]$. With the latter channel gains and power allocation for hybrid precoding, and $\B{\Psi}_i[\ell]={\B{\Upsilon}_i^{-1}[\ell]{\B{\Lambda}}_i[\ell]{\B{\Gamma}}_i[\ell]}$, the hybrid effective precoder results in
\begin{align}
\label{eq:hybridEffPrec}
\B{P}_{\text{eff},i}&=\blockdiag(\B{P}_A\B{B}_i[1]\B{\Psi}_i[1],\ldots,\B{P}_A\B{B}_i[L]\B{\Psi}_i[L]) \nonumber \\
&=\blockdiag(\B{P}_A\B{P}_\text{D}[1],\ldots,\B{P}_A\B{P}_\text{D}[L]),
\end{align}
where $\B{P}_{\text{D}}[\ell]$ contains the $d_i[\ell]$ precoders allocated for subcarrier $\ell$. Using the mappings $\pi(j)$ and $\beta_j[\ell]$ for all $j\leq i$, the precoders for all users $\B{P}_{\text{D},k}[\ell]$ are constructed with the columns of $\B{P}_\text{D}[\ell]$.

The hybrid combiners are directly obtained by projecting the vectors ${\B{g}}_i[\ell]$ onto the feasible set at each iteration \cite{StUtJoLu17}
\begin{equation}
\label{eq:hybridComb}
	\B{g}_i[\ell]=\beta_{i}[\ell]\frac{\B{S}_{\pi(i),i}\B{g}_{\text{A},i}}{\|\B{S}_{\pi(i),i}\B{g}_{\text{A},i}\|_2},
\end{equation} 
where $\B{g}_{\text{A},i}$ is obtained following the same procedure employed for the precoding matrix in \eqref{eq:HybidProj}. 
The composite channel and subsequent computations include these updates. The combining matrices for each user are next built using the mapping function and the frequency selective scalars $\beta_j[\ell]$, as stated for the precoders.   

Wideband H-\ac{LISA} procedure is summarized in Alg. \ref{alg:WHLISA}

\begin{algorithm}
	\caption{Wideband H-LISA}
	\label{alg:WHLISA}
	\begin{algorithmic}[1]
		\STATE Initialize: ${\B{Q}}^{\text{red}}_0[\ell]= [\,]$,  ${\B{H}}^{\text{red}}_{\text{comp},0}[\ell]= [\,]$, $\B{T}_1=\mathbf{I}_N$, $\B{S}_{k,1}=\mathbf{I}_R$, $R_{\text{sum},0}=0$, $d_1[\ell]=1$, $i = 0$
		\REPEAT 
		\STATE $i = i+1$
		\FORALL {$k\in\{1,\ldots,K\}$} 
		\STATE  $\B{H}_{k,i}=[\B{S}_{k,i}\B{H}_k[1]\B{T}_i,\ldots,\B{S}_{k,i}\B{H}_k[L]\B{T}_i]$
		\STATE $\B{g}_{i}(k)=\B{u}^\text{max}_{k,i}(\B{H}_{k,i})$
		\STATE ${\B{q}_{i}(k)=\B{v}_{k,i}^\text{max}\left({\left(\B{g}_{i}(k)^\He \B{H}_{k,i}\right)^\text{T-block}}\right)}$
		\STATE $\B{\mu}_{k,i}=\left(\B{g}_{i}^\He(k) \B{H}_{k,i}\right)^\text{T-block} \B{q}_{i}(k)$
		\ENDFOR
		\STATE $\pi(i)=\argmax_{k\in\{1,\ldots,K\}}\|\B{\mu}_{k,i}\|_1$
		\STATE $\B{q}_i=\B{q}_{i}(\pi(i))$, $\B{g}_i=\B{g}_{i}(\pi(i))$
		\STATE $\B{g}_i\leftarrow$ hybrid projection \eqref{eq:hybridComb}
		\FORALL {$\ell\in\{1,\ldots,L\}$} 
		\STATE ${\B{L}}_i[\ell]=\left[\begin{array}{c} {\B{H}^{\text{red}}_{\text{comp},i-1}[\ell]} \\ \B{g}_i^\He\B{H}_{\pi(i)}[\ell]
		\end{array}\right][{\B{Q}^{\text{red}}_{i-1}[\ell]},\B{q}_i]$
		\STATE ${\B{\Lambda}}_i[\ell]=[\diag({\B{L}}_i^{-\He}[\ell]{\B{L}}_i^{-1}[\ell])]^{-1/2}$
		\ENDFOR
		\STATE ${\B{\Gamma}}_i\leftarrow$ waterfilling with ${\B{\Lambda}}_i[1],\ldots,{\B{\Lambda}}_i[L]$
		\STATE $R_{\text{sum},i}\leftarrow$ compute metric \eqref{eq:sum_rate}
		\IF {$R_{\text{sum},i}>R_{\text{sum},i-1}$}
		\FORALL {$l\in\{1,\ldots,L\}$} 
		\IF {$[{\B{\Gamma}}_i[\ell]]_{d_i[\ell],d_i[\ell]}>0$} 
		\STATE ${\B{H}}^{\text{red}}_{\text{comp},i}[\ell]=\left[\begin{array}{c} \B{H}^{\text{red}}_{\text{comp},i-1}[\ell] \\ \B{g}_i^\He\B{H}_{\pi(i)}[\ell\end{array}\right]$ 
		\STATE ${\B{Q}}_{i}^{\text{red}}[\ell]=[{\B{Q}}_{i-1}^{\text{red}}[\ell],\B{q}_i]$
		\STATE  ${\B{P}}_{\text{eff},i}[\ell]={\B{Q}}_i^{\text{red}}[\ell]{\B{L}}_i^{-1}[\ell]{\B{\Lambda}}_i[\ell]{\B{\Gamma}}_i[\ell]$ 
		\STATE $d_{i+1}[\ell]=d_{i}[\ell]+1$
		\ENDIF
		\ENDFOR
		\STATE $\B{T}_{i+1}\leftarrow$ update projector with \eqref{eq:projUpdate}
		\STATE $\B{S}_{\pi(i),i+1}=\B{S}_{\pi(i),i}-\B{g}_i\B{g}_i^\He$					
		\ELSE 
		\STATE break
		\ENDIF
		\UNTIL{$i = \LBS$}
		\STATE $\B{P}_{\text{eff},i}\leftarrow$ compute hybrid precoders \eqref{eq:hybridEffPrec}
	\end{algorithmic}
\end{algorithm}

\subsection{Computational Complexity}
\label{sec:redCompl}
In this Section, we focus on the operations with larger computationally complexity of Alg. \ref{alg:WHLISA}, namely, the user selection of the first stage, the inversion of the triangular matrix in the second stage, and the projector's update. 

Recall that we look for the best candidate to allocate each of the streams. For user $k$, the computational complexities are $\mathcal{O}(R^2LN)$  to determine $\B{g}_{i}(k)$, and $\mathcal{O}(\min\{L^2N,LN^2\})$ to calculate $\B{q}_{i}(k)$. The complexity of the inverse for subcarrier $\ell$, and the update of projector $\B{T}_{i}$ are about $\mathcal{O}(d_i^{2.4}[\ell])$, and $\mathcal{O}(\rank^{2.4}(\B{H}_{k}))$, respectively, where
\begin{equation}
\label{eq:allFreqChann}
	\B{H}_{k}=[\B{H}_k^\Tr[1],\ldots,\B{H}_k^\Tr[L]]^\Tr\in\mathbb{C}^{RL\times N}.
\end{equation} 
Since $d_i[\ell]\leq \LBS$, the inverse of the triangular matrix is computationally inexpensive. Moreover, in the following section, we show empirically that the effective rank of $\B{H}_{k}$  is small compared to its size, even taking the beam squint effect into account. This behavior is related to the channel model in Sec. \ref{subsec:channel}. Observe also that the rank does not depend on the number of subcarriers. 

\begin{figure}[t]
	\centering
	\includegraphics[width=.99\columnwidth]{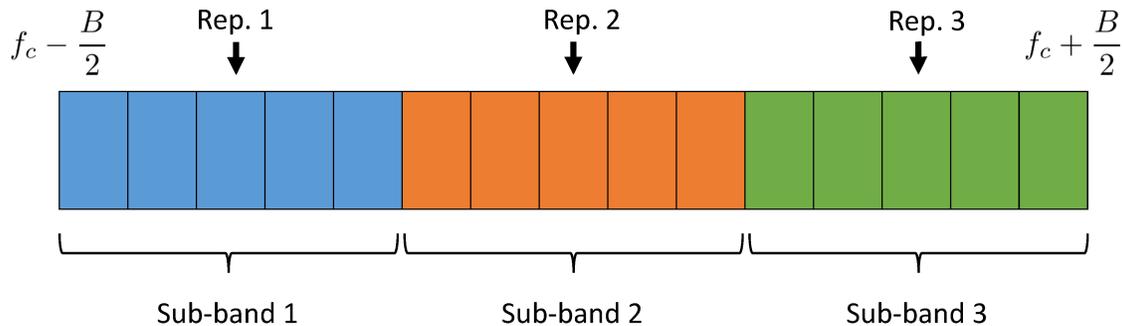}
	\caption{Example of subbands and representation subcarriers for $L=15$ subcarriers and $L_s=3$ subbands}
	\label{fig:subbands}
\end{figure}

The computational complexity of the user selection depends on the number of subcarriers, which can be large. We propose to reduce this complexity by defining subbands and a representation subcarrier for each of them, as shown in Fig. \ref{fig:subbands}. Consider for simplicity that $L$ is divisible by $2L_s$. Thus, we define the index $\ell_n=\frac{L}{2L_s}+(n-1)\frac{L}{L_s}$ with $n\in\{1,\ldots,L_s\}$. The matrices \eqref{eq:projectedChannels} and \eqref{eq:equivFrecSelChannelMatrix} are rewritten accordingly as
\begin{align}
	\B{H}_{k,i}=[\B{S}_{k,i}\B{H}_k[\ell_1]\B{T}_i,\ldots,\B{S}_{k,i}\B{H}_k[\ell_{L_s}]\B{T}_i],\\
	\left(\B{g}_{i}(k)^\He \B{H}_{k,i}\right)^\text{T-block}=\begin{bmatrix}
	\B{g}_{i}^\He(k)\B{S}_{k,i}\B{H}_k[\ell_1]\B{T}_i\\\vdots\\ \B{g}_{i}^\He(k)\B{S}_{k,i}\B{H}_k[\ell_{L_s}]\B{T}_i
	\end{bmatrix}.
\end{align}
When the number of subbands $L_s$ is chosen to satisfy $L_s\ll L$ and $L_s\ll N$, we achieve the lower complexities $\mathcal{O}(R^2L_sN)$ and  $\mathcal{O}(L_s^2N)$. Numerical results in the subsequent section reveal the feasibility of this approximation. 

\section{Simulation Results}
In this Section, we present the results of  numerical experiments to evaluate the performance of the proposed scheduling, precoding, and combining designs.

The setup considered consists of a \ac{BS} equipped with $N=64$ transmit antennas and $K=4$ users with $R=16$ antennas each. The number of channel paths	$\Npk=4$ for all users, and the number of \ac{RF} chains are $\LBS=4$ and $\Lrk=2$, respectively. The results are averaged over $1000$ channel realizations generated according to the channel model in Sec. \ref{subsec:channel}. To that end, we consider a central carrier frequency of $f_c=28$ GHz and  signal bandwidths $B=400$ MHz, $B=800$ MHz \cite{SuMaRa17}, and $B=3200$ MHz. The number of subcarriers is set to $L=32$. 

\begin{figure}[t]
	\centering
	\includegraphics[width=.99\columnwidth]{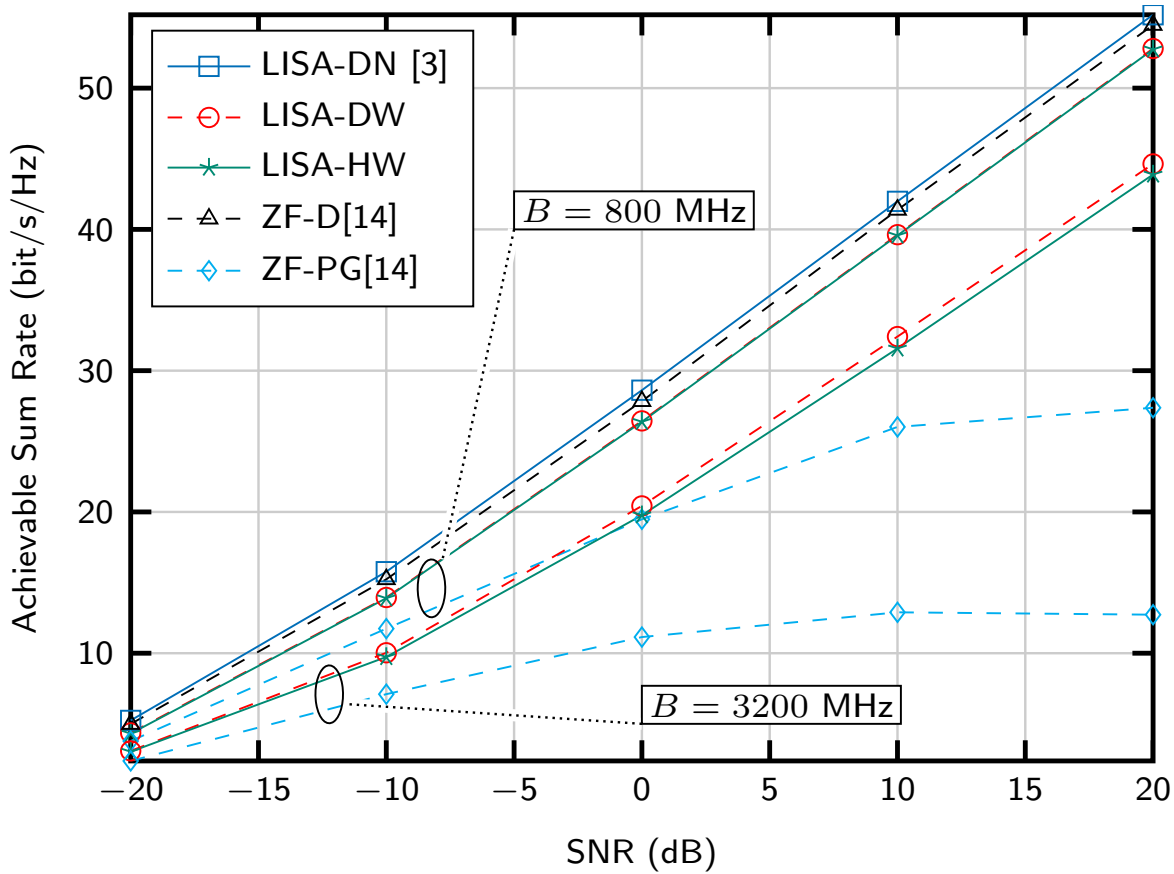}
	\caption{Numerical results for $f_c=28$ GHz and bandwidths $B=800,3200$ MHz: Sum Rate vs SNR for $K=4$ users, $N=64$ transmit antennas, $R=16$ receive antennas, $L=32$ subcarriers and $\Npk=4$ propagation paths for each user. The number of RF chains are $\LBS=4$ and $\Lrk=2$.}
	\label{fig:sumRate}
\end{figure}

In Fig. \ref{fig:sumRate}, we plot the achievable sum rates obtained with different strategies and bandwidths $B=800$ MHz and $B=3200$ MHz. We take as a benchmark the results obtained with the \ac{LISA} scheme applied individually at each subcarrier \cite{UtStJoLu18}. This strategy allows us to allocate a maximum of $\LBS$ streams at each subcarrier, though neglecting the rank constraint imposed by the frequency flat analog \ac{PS} network. Furthermore, we establish a per-subcarrier power constraint $P'_\text{tx}=\frac{P_\text{tx}}{L}$, with $P_\text{tx}$ the available total transmit power for the other strategies. This bound is labeled as \emph{LISA Digital Narrow} (LISA-DN). 
The \emph{LISA Digital wideband} (LISA-DW) and \emph{LISA Hybrid wideband} (LISA-HW) curves show the results obtained with the strategies proposed in this work. The label \emph{Digital} refers to performance results obtained with precoders and combiners whose entries are not restricted to be unit modulus. That is, Alg. \ref{alg:WHLISA} is employed but without the projections explained in Sec. \ref{sec:WHLISA}. Notice that the rank restriction imposed by the number of \ac{RF} chains $\LBS$ holds. For the \emph{Hybrid} counterpart, variable \acp{PS} with infinite resolution have been considered. Remarkably, the gap between the proposed strategies and the \emph{Digital narrow} approximation is constant for the high SNR regime, while the performance loss due to the hybrid approximation is negligible. Moreover, we compare to the results obtained with the \emph{Digital} and hybrid \emph{Projected Gradient} zero-forcing methods in \cite{GoRoGoCaHe18} (ZF-D) and (ZF-PG), where common support was assumed for the different subcarriers. Therein, authors imposed power constraints for each user and subcarrier, leading to $P'_\text{tx}=\frac{P_\text{tx}}{LK}$ to provide a fair comparison. As a result of the beam squint effect and the hybrid decomposition inaccuracy, the gap with respect to the wideband strategies proposed in this work increases with the SNR. This effect comes from the fact that the strategy assuming common support decomposes a digital design into its analog and digital baseband counterparts. Accordingly, when the number of utilized  spatial dimensions increases in the high SNR regime, the accuracy of the hybrid decomposition reduces due to the lack of similarity of the channel responses for different subcarriers. This is directly related to the ratio $\frac{\xi[\ell]}{f_c}$ and to the number of channel paths $\Npk$. Consequently, the performance loss compared to the proposed benchmark is greater for $B=3200$ MHz. Similar gaps are obtained for equal relative bandwidths, e.g., $f_c=56$ GHz and $B=1600$ MHz, or $f_c=84$ GHz and $B=2400$ MHz, corresponding to $f_c=28$ GHz and $B=800$ MHz.

\begin{figure}[t]
	\centering
	\includegraphics[width=.99\columnwidth]{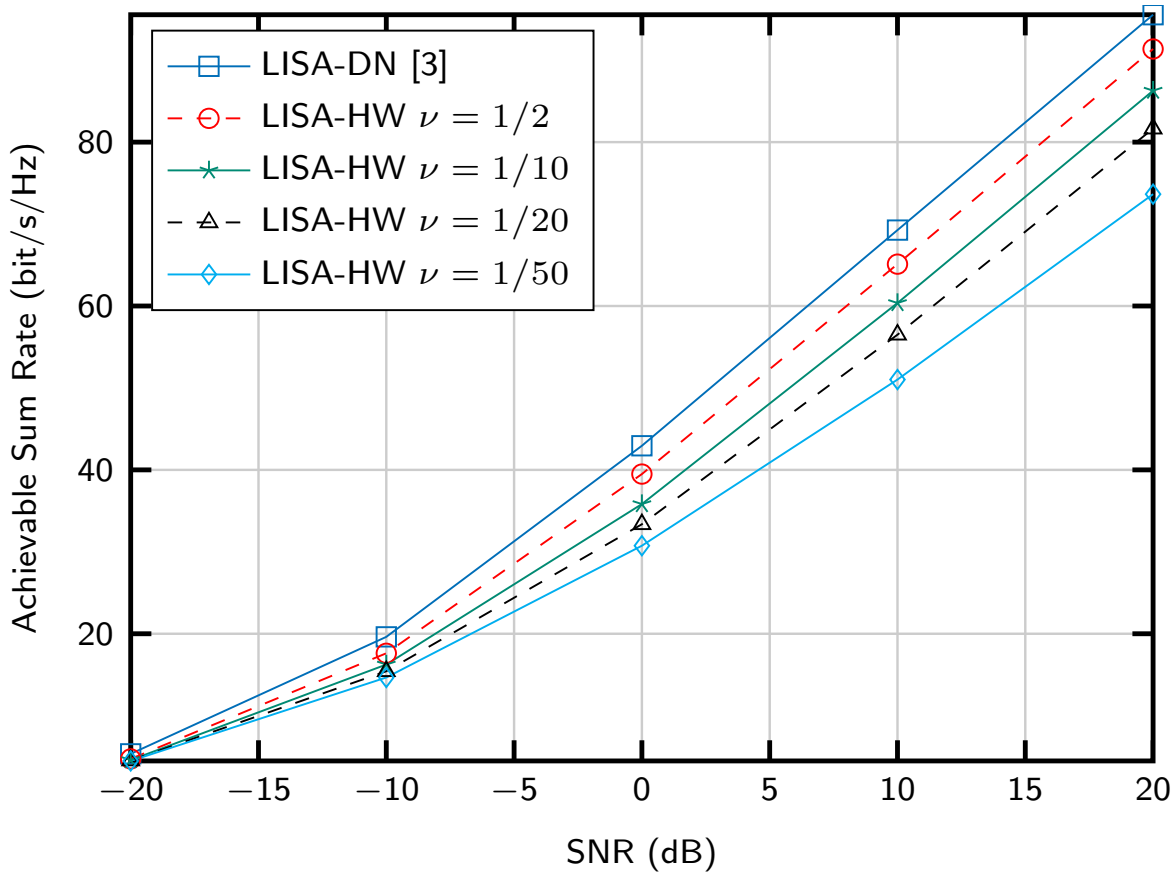}
	\caption{Numerical results for carrier frequency $f_c=28$ GHz and signal bandwidth $B=400$ MHz: Sum Rate vs SNR for the relaxed interference constraint and different thresholds $\nu=\frac{1}{2},\frac{1}{10},\frac{1}{20},\frac{1}{50}$. $K=4$ users, $N=64$ transmit antennas, $R=16$ receive antennas, $L=32$ subcarriers and $\Npk=4$ propagation paths. The number of RF chains are $\LBS=8$ and $\Lrk=2$.}
	\label{fig:slope}
\end{figure}

\begin{figure}[t]
	\centering
	\includegraphics[width=.99\columnwidth]{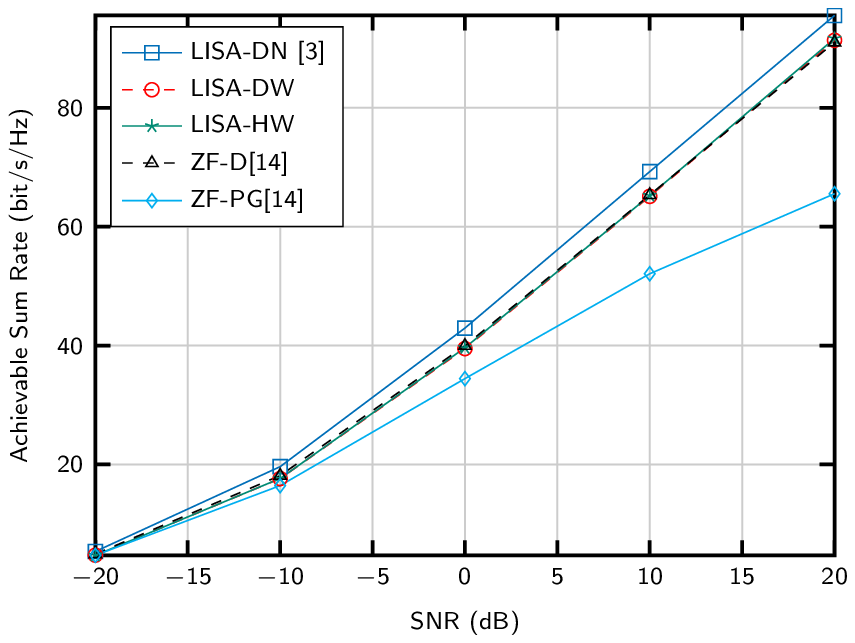}
	\caption{Numerical results for carrier frequency $f_c=28$ GHz and signal bandwidth $B=400$ MHz: Sum Rate vs SNR $K=4$ users, $N=64$ transmit antennas, $R=16$ receive antennas, $L=32$ subcarriers and $\Npk=4$ propagation paths for each user. The number of RF chains are $\LBS=8$ and $\Lrk=2$.}
	\label{fig:sumRateCommon}
\end{figure}

We evaluate the former strategies in a scenario where the relative signal bandwidth is small, namely $B=400$ MHz for $f_c=28$ GHz, and the channel model of Sec. \ref{subsec:channel} can be approximated by setting   $\B{\Xi}_{\text{MS},k}(\theta_{k,p})[\ell]=\mathbf{I}_R$ and $\B{\Xi}_{\text{BS},k}(\phi_{k,p})[\ell]=\mathbf{I}_N$. This setup is similar to that in \cite{GoRoGoCaHe18} and the references therein. Moreover, we set the number of available \ac{RF} chains at the \ac{BS} to $\LBS=8$. The throughput curves for relative small signal bandwidth are plotted in Fig. \ref{fig:slope} and Fig. \ref{fig:sumRateCommon}. 
Larger number of \ac{RF} chains make it possible to allocate more streams, but obviously the proposed algorithm quickly runs out of degrees of freedom on the feasible subspace for precoding as described in Section \ref{sec:FistStage}, which leads to a reduced slope of the achievable rate curve. In order to mitigate this undesirable effect, the zero interference constraint in the first stage of Alg. \ref{alg:WHLISA} is relaxed, thus allowing for a certain level of interference. This can be achieved by considering the significant singular vectors from the set $\{\B{T}_i\B{H}_{\pi(i)}^\He[\ell]{\B{g}_{i}[\ell]}\}_{\ell=1}^L$, according to a certain threshold $\nu$, in the projector's update of \eqref{eq:projUpdate}. Consequently, the constraints in    
\eqref{eq:widebandLISAproduct} will only be met approximately. This rank reduction in the projector means that the zero-forcing condition  strongly depends on the second stage of Alg. \ref{alg:WHLISA}, i.e., on the baseband precoders for each subcarrier. We observe in our experiments that for the high SNR regime this approach leads to diagonally dominant matrix products in \eqref{eq:widebandLISAproduct}.
The performance results for different thresholds $\nu$ are shown in Fig. \ref{fig:slope}, where we just considered the singular values larger than $\sigma_\text{max}\nu$ to update the projector, with $\sigma_\text{max}$ being the dominant singular value. Notice that for the limit case $\B{\Xi}_{\text{MS},k}(\theta_{k,p})[\ell]=\mathbf{I}_R$ and $\B{\Xi}_{\text{BS},k}(\phi_{k,p})[\ell]=\mathbf{I}_N$ the interference for all subcarriers of user $k$ lie in the common subspace spanned by  $\Npk$ steering vectors. On the contrary, when beam squint is considered the interference is not restricted to this particular subspace. As a consequence, this interference relaxation does not increase the number of allocated streams when the beam squint effect is strong.	
Fig. \ref{fig:sumRateCommon} exhibits the good performance achieved with Alg. \ref{alg:WHLISA} considering the interference relaxation at the first stage with threshold $\nu=\frac{1}{2}$. In addition, the  method in \cite{GoRoGoCaHe18} allocates $2$ streams to each user and achieves better results in this particular scenario. Nevertheless, since the hybrid approximation of the digital design is based on a steepest descent method computationally complexity might be high.

\begin{figure}[t]
	\includegraphics[width=.99\columnwidth]{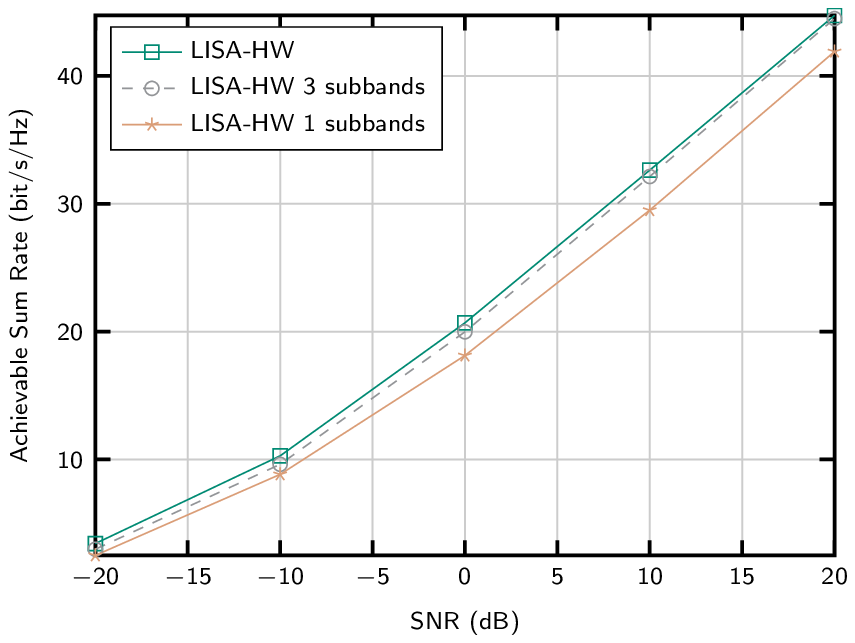}
	\caption{Sum Rate vs SNR considering $f_c=28$ GHz and $B=3200$ MHz. We use $L_s=3$ and $L_s=1$ subbands for $K=4$ users, $N=64$ transmit antennas, $R=16$ receive antennas, $L=32$ subcarriers and $\Npk=4$ propagation paths. The number of RF chains are $\LBS=4$ and $\Lrk=2$. }
	\label{fig:LISA3b}
\end{figure}

In section \ref{sec:redCompl}, we proposed to reduce the computational complexity of the user selection performed at the first stage of \ac{LISA} (see Sec. \ref{sec:FistStage}). The applicability of the approximation based on subbands is shown in Fig. \ref{fig:LISA3b}. We set the number of subbands to $L_s=3$ and $L_s=1$, and compare the performance achieved with the hybrid precoders and combiners with that obtained using the information of all subcarriers, considering $f_c=28$ GHz and $B=3200$ MHz. From numerical results we observe performance losses about $36\%$ and $17\%$, for $L_s=1$ and $L_s=3$ and $\text{SNR}=-20$ dB. The relative losses reduce to $18\%$  and $9\%$, respectively, for $\text{SNR}=-10$ dB. In the high SNR regime the approach considering $L_s=3$ subbands suffices to achieve a very good performance, while the computational complexity reduction is significant. 

The following experiment highlights the effects of signal bandwidth and carrier frequency on the average effective rank for the channel matrix comprising all subcarriers, $\B{H}_k$ in \eqref{eq:allFreqChann}. The $n$-th singular value $\sigma_{k,n}$ of matrix $\B{H}_k$ was assumed to be non-negligible if $\sigma_{k,n}\geq \frac{\sigma_{k,1}}{50}$, with $\sigma_{k,1}\geq\sigma_{k,2}\geq\ldots\geq\sigma_{k,N}$. Thus, we averaged the number of non-negligible singular values over $1000$ channel realizations.  The results are shown in Table \ref{tab:channelEffRank} and exhibit the small average effective rank for $\B{H}_k$, compared to the matrix size  $16\cdot32\times 64$. Accordingly, a small set of vectors is enough to compute the orthogonal projector $\B{\Pi}_i^\perp$ of \eqref{eq:projUpdate}.
\setcounter{table}{-1}
\setlength{\tabcolsep}{0.3em}
\renewcommand*{\arraystretch}{1.5}
\setlength{\tabcolsep}{0.3em}
\renewcommand*{\arraystretch}{1.5}
\begin{table}[h]
	\begin{footnotesize}
		\begin{center}
			\begin{tabular}{|c|c|c|c|}
				\hline
				$\Npk$&$f_c$&$B$& Avg. Eff. Rank\\
				\hline \hline
				$4$&$28$GHz&$400$MHz&$7.2$\\\hline
				$4$&$28$GHz&$800$MHz&$8.9$\\\hline
				$4$&$28$GHz&$3200$MHz&$15.2$\\\hline
				$4$&$60$GHz&$400$MHz&$6.5$\\\hline
				$4$&$60$GHz&$800$MHz&$7.1$\\\hline
				$4$&$60$GHz&$3200$MHz&$10.98$\\\hline
			\end{tabular}
			\caption{Channel effective ranks}
		\end{center}
	\end{footnotesize}
	\caption{Average effective channel ranks for different setups with $N=64$ and $R=16$.}
	\label{tab:channelEffRank}
\end{table}

\begin{figure}[t]
	\includegraphics[width=.99\columnwidth]{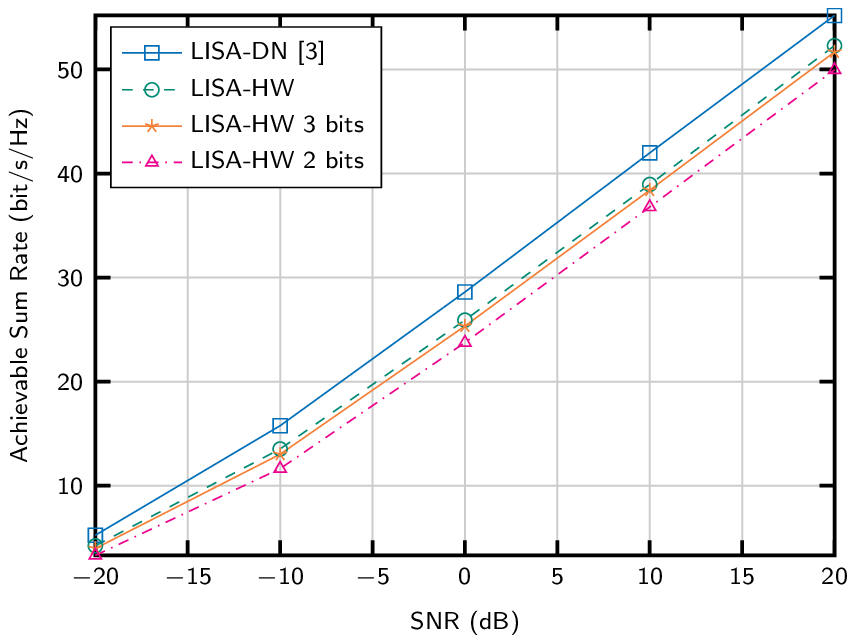}
	\caption{Sum Rate vs SNR using variable \acp{PS} with $2$ and $3$ resolution bits, for $K=4$ users, $N=64$ transmit antennas, $R=16$ receive antennas, $L=32$ subcarriers and $\Npk=4$ propagation paths for each user. The number of RF chains are $\LBS=4$ and $\Lrk=2$.}
	\label{fig:LISAQuant}
\end{figure}

Thus far we have considered infinite resolution variable \acp{PS} to implement the analog precoders and combiners.  In Fig. \ref{fig:LISAQuant}, we evaluate the performance of the proposed method for \ac{PS} resolutions of $3$ and $2$ bits. The carrier frequency is $f_c=28$ GHz for a bandwidth of $B=800$ MHz. Results show that the performance loss due to quantization is negligible for \ac{PS} with $2^3=8$ available phases and moderate for \ac{PS} with only $2^2$ quantized values.

\begin{figure}[t]
	\includegraphics[width=.99\columnwidth]{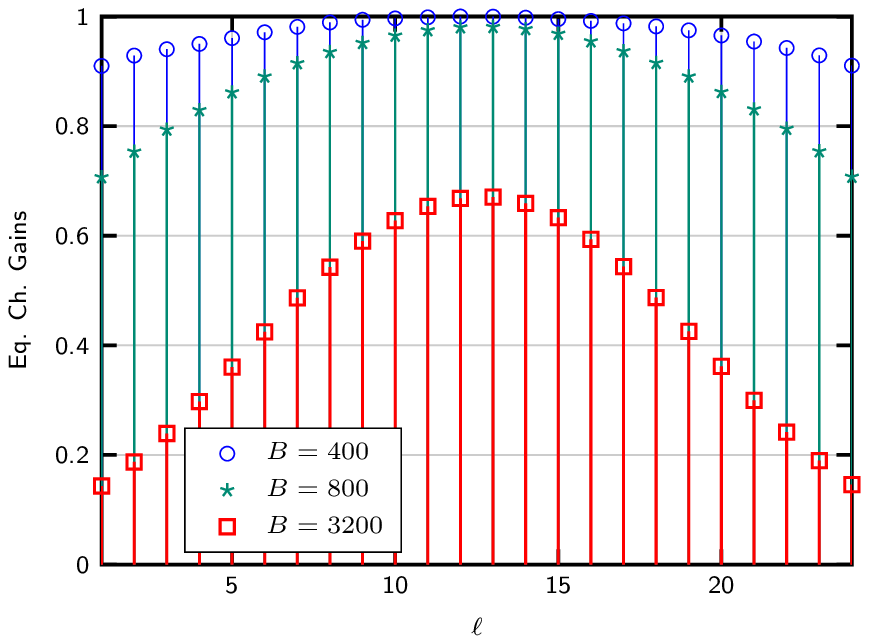}
	\caption{Avg. equivalent channel gains vs subcarrier index, for $K=4$ users, $N=64$ transmit antennas, $R=16$ receive antennas, $L=24$ subcarriers and $\Npk=4$ propagation paths for each user. The number of RF chains are $\LBS=4$ and $\Lrk=2$, with carrier frequency  $f_c=28$ GHz and signal bandwidths $B=400,800,3200$ MHz.}
	\label{fig:eqChGains}
\end{figure}

\begin{figure}[t]
	\includegraphics[width=.99\columnwidth]{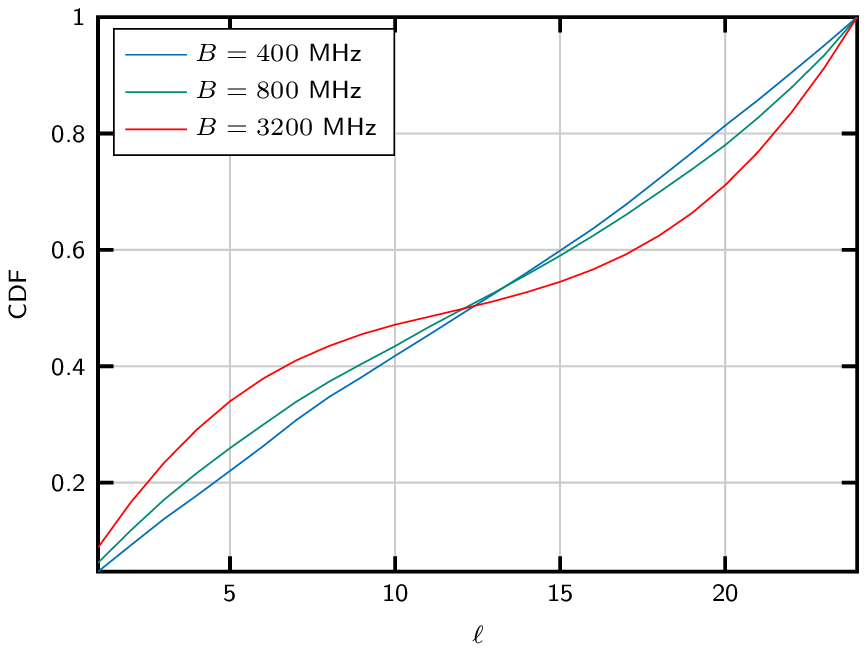}
	\caption{Empirical conditional \ac{CDF} of a switched-off subcarrier with respect to the subcarrier index $\ell=1,\dots,L$ for SNR$=-20$ dB.  $K=4$ users, $N=64$ transmit antennas, $R=16$ receive antennas, $L=24$ subcarriers and $\Npk=4$ propagation paths for each user. The number of RF chains are $\LBS=4$ and $\Lrk=2$, with carrier frequency  $f_c=28$ GHz and signal bandwidths $B=400,800,3200$ MHz.}
	\label{fig:ecdf}
\end{figure}

To get a better understanding of the impact of beam squint over Alg. \ref{alg:WHLISA}, we plot in Fig. \ref{fig:eqChGains} the average normalized equivalent channel gains, $\lambda_j^2[\ell]$ of \eqref{eq:perStreamRate}, for each subcarrier. We present a comparison for $f_c=28$ GHz and different signal bandwidths $B$, which clearly reveals the gain losses incurred due to beam squint. The gain distribution among the subcarriers observed in the figure comes from the selection of the auxiliary combiners and precoders in \eqref{eq:gaux} and \eqref{eq:qaux}, which aim at finding the largest gains jointly for all the subcarriers. The selection apparently prefers center frequencies in order to capture the whole bandwidth as much as possible by means of a flat solution.
Moreover, these losses exhibit similar behavior irrespective of the SNR regime. 
However, the power allocation is determined according to the values of $\lambda_j^2[\ell]$ and SNR. For the low SNR regime, the waterfilling power allocation selects the largest gain candidates. Therefore, the subcarriers close to the central frequency receive more power, whereas the subcarriers in the edges allocate a smaller portion of the power budget. 
When the allocated power for a particular subcarrier $\ell$ is zero, this subcarrier is switched off (the frequency selective scalar $\beta_i[\ell]$ is set to $0$ in Alg. \ref{alg:WHLISA}).  Also, the number of subcarriers with zero power increases for large bandwidths, according to this unfair power allocation. On the other hand, for the high SNR regime the waterfilling allocation evenly distributes available transmit power among the subcarriers. Therefore, the unfair power sharing for different frequencies vanishes as the SNR increases. Indeed, the probability of switching off subcarriers in the high SNR regime is very small. Thus, given a switched-off subcarrier, we evaluate the conditional frequency that this subcarrier occupies a particular carrier index $\ell\in\{1,\ldots,L\}$. Recall that index $\ell$ is related to the carrier frequency through $f[\ell]$ (see Sec. \ref{subsec:channel}). Accordingly, Fig. \ref{fig:ecdf} illustrates the empirical conditional \ac{CDF} for different signal bandwidths and the carrier frequency $f_c=28$ GHz. Whereas the position of switched-off subcarriers is almost uniformly distributed for small bandwidth, in case of larger bandwidth, the likelihood of switching off a subcarrier obviously increases when it is close to the edge frequencies. This effect is accentuated for $B=3200$ MHz and virtually disappears when considering $B=400$ MHz.

\section{Conclusion}
This work jointly addresses user scheduling and hybrid precoding and combining designs for a wideband multiuser \ac{mmWave} communications system. The main limitation of the hybrid architecture is that the analog precoder has to be jointly designed for all users and subcarriers. Similarly, the analog combiner is common for all subcarriers and a particular user. To circumvent this difficulty, we propose to employ the information of all the subcarriers to allocate a data stream to the best user candidate. Moreover, the proposed method provides the additional flexibility of switching off subcarriers. The following stage removes the remaining inter-stream interference for each subcarrier using the frequency selective digital precoders and determines the power allocation.  The proposed method exhibits excellent performance in the numerical experiments, and is particularly suitable to overcome the so-called beam squint effect.    

\appendix
\subsection{Combiners Linear Dependence}
\label{ap:Sproj}
	At the $i$-th iteration of \ac{LISA}, we assume that $\pi(i)=\pi(j)$ for any $j$ such that $j<i$. Consider that ${\beta}_i[\ell]\neq 0$ and ${\beta}_j[\ell]\neq 0$ for the $\ell$-th subcarrier, such that $\B{g}_i[\ell]=\B{g}_i$ and $\B{g}_j[\ell]=\B{g}_j$ with $\B{g}_i^\He\B{g}_j\neq 0$, and the subsequent decomposition $\B{g}_i=\alpha\B{g}_j+\nu\B{g}_j^\perp$ where $\alpha=\B{g}_j^\He\B{g}_i$ and $\nu=\|\B{g}_i-\alpha\B{g}_j\|_2$. The resulting product of the linear dependent component with the projected channel for the $\ell$-th subcarrier reads as
	\begin{equation}
	\alpha^*\B{g}_j^\He\B{H}_{\pi(j)}[\ell]\B{T}_i=\alpha^*\B{g}_j^\He\B{H}_{\pi(j)}[\ell]\B{T}_j\B{\Pi}_{j}^\perp\cdots\B{\Pi}_{i-1}^\perp=\mathbf{0}^\text{T}\notag\\
	\end{equation}
	since $\B{\Pi}_{j}^\perp$ is the projector onto  $\text{null}\{\B{g}_j^\He\B{H}_{\pi(j)}[\ell]\B{T}_j\}_{\ell=1}^L$.

 \newpage
 \bibliographystyle{IEEEtran}
 %
 \bibliography{references}
\end{document}